\documentclass[10pt,final,journal,twocolumn]{IEEEtran}
\ifCLASSINFOpdf
\else
\fi
\usepackage{cite}
\usepackage[cmex10]{amsmath}
\usepackage{amsthm}
\usepackage{algorithmic}
\usepackage{stfloats}
\usepackage{multicol,multienum}
\usepackage[dvips]{graphicx}
\graphicspath{{./figurespdf/}}
\usepackage{algorithm} 
\usepackage{algorithmic} 
\usepackage{multirow} 
\usepackage{amsmath}
\usepackage{xcolor}
\usepackage{subfigure}

\newtheorem{theorem}{\textbf{Theorem}}

\newtheorem{definition}{\textbf{Definition}}

\usepackage{amssymb}

\begin{document}

\title{Distributed Relay Selection for Heterogeneous UAV Communication Networks Using A Many-to-Many Matching Game Without Substitutability}

%

\author{\IEEEauthorblockN{Dianxiong Liu\IEEEauthorrefmark{1}\IEEEauthorrefmark{2},
Yuhua Xu\IEEEauthorrefmark{1}\IEEEauthorrefmark{2},
Yitao Xu\IEEEauthorrefmark{1},
Qihui Wu\IEEEauthorrefmark{3},
Jianjun Jing\IEEEauthorrefmark{1},
Yuanhui Zhang\IEEEauthorrefmark{4} and
Alagan Anpalagan\IEEEauthorrefmark{5}}

\IEEEauthorblockA{\IEEEauthorrefmark{1}College of Communications Engineering, PLA Army Engineering University, Nanjing, China. \\\IEEEauthorrefmark{2}Science and Technology on Communication Networks Laboratory, Shijiazhuang, China.\\\IEEEauthorrefmark{3}College of Electronic and Information Engineer, Nanjing University of Aeronautics and Astronautics, Nanjing, China.\\\IEEEauthorrefmark{4}PLA Zhenjiang Watercraft College, China. \\ \IEEEauthorrefmark{5}Department of Electrical and Computer Engineering, Ryerson University, Toronto, Canada. \\Emails: dianxiongliu@163.com, yuhuaenator@gmail.com, yitaoxu@126.com, wuqihui2014@sina.com, jianjun.jing@panda.cn, zyh0126@foxmail.com, alagan@ee.ryerson.ca}
}

\IEEEpeerreviewmaketitle
\maketitle

\begin{abstract}
This paper proposes a distributed multiple relay selection scheme to maximize the satisfaction experiences of unmanned aerial vehicles (UAV) communication networks. The multi-radio and multi-channel (MRMC) UAV communication system is considered in this paper. One source UAV can select one or more relay radios, and each relay radio can be shared by multiple source UAVs equally. Without the center controller, source UAVs with heterogeneous requirements compete for channels dominated by relay radios. In order to optimize the global satisfaction performance, we model the UAV communication network as a many-to-many matching market without substitutability. We design a potential matching approach to address the optimization problem, in which the optimizing of local matching process will lead to the improvement of global matching results. Simulation results show that the proposed distributed matching approach yields good matching performance of satisfaction, which is close to the global optimum result. Moreover, the many-to-many potential matching approach outperforms existing schemes sufficiently in terms of global satisfaction within a reasonable convergence time.
\end{abstract}

\begin{IEEEkeywords}
UAV communication networks, Distributed multiple relay selection, Satisfaction experiences, Many-to-many matching market without substitutability, Potential matching.
\end{IEEEkeywords}

\section{Introduction}
UAV communications can improve the breadth and dimension of wireless communication\cite{UAVcommunication}. Large-scale model of UAVs is the developing trend, and it is an urgent problem to solve the communication problem among the group of UAVs. There is a cluster head in the UAV group, keeping communication with the ground control center and communicating to other UAVs. Each UAV needs to maintain the communication connection with the UAV controller. However, due to the limitation of transmit power, some UAVs can not communicate directly with the controller. Thus, the relay technology is needed in UAV communication networks\cite{UAVcommunication}.

In the relay model with a large number of UAVs, only suitable relay selections can lead to better performances. Although the relay selection of ground communication has been studied \cite{Nodirect,IET,multione,multitwo,multithree,multitomulti1,multitomulti3}, these results are not fully applicable to UAV networks. The reason is that one UAV may need to complete diverse information transmissions for different tasks at the same time. Therefore, multiple radio systems equipped to UAV networks are available such as \cite{multitomulti1,multitomulti3}. Thus, it is meaningful to propose selection strategies of multiple nodes. However, previous existing works mainly focus on problems of one-to-one relay selection strategies \cite{Nodirect,IET} or many-to-one relay selection strategies \cite{multione,multitwo,multithree}. These selection approaches are not suitable to many-to-many UAV communication systems, in which strategies of relay selection will be more complex.

Moreover, existing literatures \cite{Nodirect,IET,multione,multitwo} mainly focus on the throughput performance, ignoring the actual demands of users. It may be inadvisable to overemphasize the throughput performance when requirements of users are heterogeneous. Therefore, this paper mainly develops more reasonable relay selections with the heterogeneous requirements of UAVs. A source UAV (SV) with large capacity requirements can connect to multiple radios of relay UAVs (RVs), while the SV with a low capacity requirement connects to less relay radios.

In complex UAV networks, centralized approaches may bring a large amount of information overheads. Therefore, it is necessary to develop distributed approaches. However, it is a difficult problem for decision makers to optimize the performance of whole network according to their own selection strategy. In this case, the UAV relay network can be modeled as a many-to-many matching market \cite{matching4}. Matching  models are powerful and promising to address the assignment problem in wireless networks \cite{matching2,matching3,matching1}. However, all the existing models above are matching markets with substitutability, in which players obtain resources by replacing the other players. These models are not suitable to model relay sharing networks without substitutability.

To solve this problem, we model the UAV relay network as a many-to-many matching game without substitutability. In the game model, SVs and RVs have their individual utilities, so as to make selection strategies respectively. SVs share the time resource of the access channel equally. According to the matching performance, RVs determine whether to accept requests of SVs or not. Different from the traditional matching game, the proposed game is non-substituted. To the best of our knowledge, it is the first application of matching market without substitutability in wireless networks.

Inspired by the potential function in \cite{potentialgame}, We propose a distributed potential matching algorithm (PMA). The performance of the global utility will be improved as the optimizing of local matching results. It shows that matching results can achieve a global stable matching of satisfaction. The performances of the proposed algorithm are evaluated in terms of satisfaction and convergence rate, which show that the proposed algorithm can improve the global satisfaction within a reasonable convergence time.

The rest of this paper is organized as follows. In Section II, the system model is provided and the relay selection problem is formulated. In Section III, the application of many-to-many matching game in satisfaction-aware relay assignment is analyzed. In Section IV, the potential matching approach for the many-to-many matching market is proposed. In Section V, the simulation results are shown and the performance of our distributed algorithm is analyzed. Finally, the conclusion is drawn in Section VI.

\section{System Model and Problem Formulation}\label{SEC:2System Model}
\subsection{System Model}

\begin{figure}
  \includegraphics[width=3in]{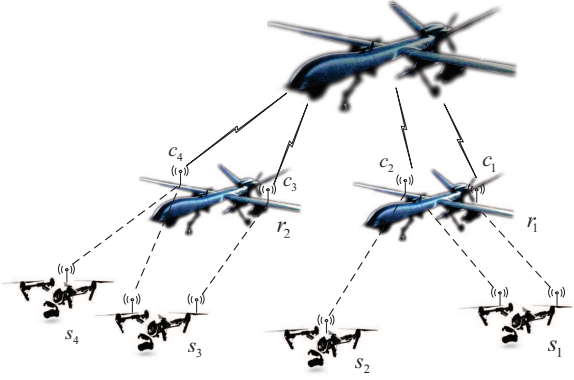}\\
  \centering
  \caption{A multi-channel and multi-radio UAV relay network.}\label{systemmodel}
\end{figure}

We consider a UAV communication network with $N$ SVs, $M$ RVs and one destination (UAV controller), where each source-destination pair $\left\{ {{s}_{n}},d \right\}$ has an opportunity to be assisted by RVs. The sets of SVs and RVs are denoted by $\mathcal{S}=\left\{ {{s}_{1}},{{s}_{2}},...,{{s}_{N}} \right\}$ and $\mathcal{R}=\left\{ {{r}_{1}},{{r}_{2}},...,{{r}_{M}} \right\}$, respectively. Each UAV is equipped with one or more radios so that each SV can connect to different radios of RVs at the same time. Radios of RVs can serve as transmitters at given channels.

The set of relay radios is denoted by $\mathcal{C}=\left\{ {{c}_{1}},{{c}_{2}},...,{{c}_{L}} \right\}$. We assume that there is no inter-channel interference because the number of orthogonal channels is enough for radios of RVs \cite{multione,multitwo}. Due to the interference constraint, there is no capacity benefit if two different radios of one node work in a same channel.

Each radio works in a half-duplex mode and the communication is frame-by-frame fashion. Each frame is divided into two time slots. The first time slot is used for the link of the SV to the RV, while the second time slot is used for the link of the RV to the destination by relay modes. There are two main relay modes, amplify-and-forward (AF) and decode-and-forward (DF) \cite{AFDF}. In this paper, we use the AF mode, and the proposed scheme can be extended to DF or hybrid modes.

Under the AF mode, RVs receive signals from SVs and then amplify and transmit them to the destination. In order to represent the performance of relay assignment among SVs, we assume that the direct transmission diversity between SVs and the destination would not exist \cite{Nodirect}. The capacity without direct transmission diversity is expressed as, \begin{equation}\label{C_AF2}
   {C_{AF}}\left( {s,r\left( c \right),d} \right) = \frac{W}{2} \cdot {\log _2}\left( {1 + \frac{{{\gamma _{sr}}{\gamma _{rd}}}}{{1 + {\gamma _{sr}} + {\gamma _{rd}}}}} \right),\
\end{equation}
where the transmitting bandwidth is denoted by $W$. The SV, radio of RV and destination are denoted by $s$, $r(c)$ and $d$, respectively. For ease of expression, the signal-to-noise-ratio (SNR) at the RV coming from the SV is denoted by ${{\gamma }_{sr}}$, ${{\gamma }_{sr}}={{{P}_{s}}{{\left| {{h}_{s,r}} \right|}^{2}}}/{{{\sigma }^{2}}}\;$. The transmission power of the SV is denoted by ${{P}_{s}}$. ${{h}_{s,r}}$ captures the path gain between the SV and the RV, and the parameter ${{\sigma }^{2}}$ is the variance of white Gaussian noise (AWGN). Similarly, the SNR from the SV and the RV to the destination are denoted by ${{\gamma }_{sd}}$ and ${{\gamma }_{rd}}$, respectively.

We assume that each radio of one SV can connect to one radio of the RV at most, so each SV can be connected to a number of relay radios that do not exceed the number of radios it equipped with. That is,
\begin{equation}\label{numberconnect}
    \sum\nolimits_{{c_l} \in {\cal C}} {{\delta _{nl}} \le {\alpha _n}},
\end{equation}
where ${{\delta }_{nl}}$ is the indicator function, which equals to 1 if SV ${{s}_{n}}$ connects to relay radio ${{c}_{l}}$ successfully, or else equals to 0 if failed. The number of radios of ${{s}_{n}}$ is denoted by ${{\alpha }_{n}}$.

In relay networks, multiple SVs will share time resources equally if they connect to a same relay radio \cite{multitwo}. The number of SVs connecting to one relay radio ${{c}_{l}}$ is denoted by $A\left( {{c}_{l}} \right)$. Therefore, obtained throughput of ${{c}_{l}}$ can be denoted by ${{B}_{l}}$. That is,
\begin{equation}\label{Relay_obtain}
    {B_l} = \frac{{\sum\nolimits_{{s_n} \in {\cal S}} {{C_{AF}}\left( {{s_n},{r_m(c_l)},d} \right){\delta _{nl}}} }}{{A\left( {{c_l}} \right)}},\
\end{equation}
where $\sum\nolimits_{{{s}_{n}}\in \mathcal{S}}{{{\delta }_{nl}}}$ equals to $A\left( {{c}_{l}} \right)$. Similarly, with the help of multiple relay radios, the data rate of ${{s}_{n}}$ can be denoted by ${{u}_{n}}$. That is,
\begin{equation}\label{source_obtain}
    {u_n} = \sum\nolimits_{{c_l} \in {\cal C}} {\frac{{{C_{AF}}\left( {{s_n},{r_m(c_l)},d} \right){\delta _{nl}}}}{{A\left( {{c_l}} \right)}}}.
\end{equation}
\begin{equation}\label{subject_to}
    {\rm{s}}{\rm{.t}}{\rm{.~~~~}}\sum\nolimits_{{c_l} \in {\cal C}} {{\delta _{nl}} \le {\alpha _n}}
\end{equation}
It should be noted that the performance of one source-relay pair would be influenced by the other SVs which have the same radio choices.

In UAV networks with heterogeneous requirements, each SV has individual type of service. Therefore, each SV aims to find the set of relay radios to meet the throughput requirement. Let ${{{u}'}_{n}}$ denote the throughput requirement of SV ${{s}_{n}}$, and the data rate is used for the utility function. Let ${{f}_{n}}\left( {{\psi }_{n}},{{\psi }_{-n}} \right)$ denote the satisfaction index of ${{s}_{n}}$, where ${{\psi }_{n}}=\left\{ {{\psi }_{n}}\left| {{\psi }_{n}}\in \mathcal{C} \right. \right\}$ is the selection strategy of ${{s}_{n}}$ and elements of $\psi $ are relay radios. ${{\psi }_{-n}}$ represents selection strategies of the other SVs. Without loss of generality, we assume that the form of the satisfaction function is a universal sigmoid function, which can describe different communication services of utility functions of SVs \cite{functionfrom2},
\begin{equation}\label{satisfaction_f}
    {f_n}\left( {{\psi _n},{\psi _{ - n}}} \right) = \frac{1}{{1 + \exp \left[ { - \lambda \left( {{u_n} - {{u'}_n} + \frac{\nu }{\lambda }} \right)} \right]}},\
\end{equation}
where we set $\nu >7$ so that ${{f}_{n}}\left( {{\psi }_{n}},{{\psi }_{-n}} \right)\ge \frac{1}{1-\exp \left( 7 \right)}\approx 1$ when the obtained throughput higher than the required throughput, i.e., ${{u}_{n}}\ge {{{u}'}_{n}}$. The satisfaction function is shown as a slightly S-shaped curve, which is suitable to express different types of demands. $\lambda $ denotes the trend-changing speed which reflects the requirement degree of communication services.

\subsection{Problem Formulation}
The problem of multiple relay selection needs to be solved to improve the global satisfaction experience. SVs choose RVs according to their throughput requirements. As shown in Fig. \ref{systemmodel}, some SVs need to access multiple relay radios simultaneously to achieve transmission rate requirements. In this case, the optimization problem of global satisfaction can be defined based on the system model,
\begin{equation}\label{global_objective}
    {\rm{maxmize~~}}\Lambda  = \sum\limits_{{s_n} \in {\cal S}} {{f_n}\left( {{\psi _n},{\psi _{ - n}}} \right)},~~
\end{equation}
where the global satisfaction utility is denoted by $\Lambda$. The proposed problem in (\ref{global_objective}) aims to maximize the aggregate satisfaction of all SVs in the network.


\section{Satisfaction-aware relay assignment as a many-to-many matching game}\label{SEC:4Algorithm}
\subsection{Many-to-many matching game model}\label{Sec:StableMatching}
Matching game theory is a powerful decentralized approach to develop the issue of resource allocation without a center controller and global information exchange \cite{matching2}. The complex resource assignment problem can be modeled as a large distributed solution by defining individual utilities for two sets of players.

In this paper, the problem of relay selection is modeled as a many-to-many matching market, which is a promising research field but it is still rarely developed. In the model, each SV ${{s}_{n}}\in \mathcal{S}$ will be assigned to one or more radios ${{c}_{l}}\in \mathcal{C}$ controlled by RVs, and each relay radio can also be shared by multiple SVs. The definition of the many-to-many matching game is given as follow \cite{matching4},

\begin{definition}\label{FuncDefinition1}
A many-to-many matching $\mu $ is a mapping by two sets of players $\left( \mathcal{S},\mathcal{C} \right)$ and two preference relations ${{\succ }_{n}},{{\succ }_{l}}$. Each player ${{s}_{n}}\in \mathcal{S}$  and ${{c}_{l}}\in \mathcal{C}$ constructs preference lists over one another, ranking the players in $\mathcal{S}$ and $\mathcal{C}$, respectively. The matching process is constrained to,
\end{definition}
\begin{itemize}
\item $\mu \left( s \right)$ is contained in $\mathcal{C}$, and $\mu \left( c \right)$ is contained in $\mathcal{S}$;
\item $\left| \mu \left( s \right) \right|\le {{q}_{s}}$ for all ${{s}_{n}}\in \mathcal{S}$;
\item $s$ is in $\mu \left( c \right)$ if and only if $c$ is in $\mu \left( s \right)$
\end{itemize}
where the preference relation $\succ $ is defined as a complete and reflexive binary relation between players in $\mathcal{S}$ and $\mathcal{C}$. $\mu \left( s \right)$ means the subset of relay radios which are connected by SV $s$, and $\mu \left( c \right)$ is the subset of SVs connecting to relay radio $c$. The quota is the maximal number of each player can match, denoted by $q$. The second constraint means that quotas of SVs are fixed while quotas of RVs are dynamic. The third constraint ensures that the matching is the mutual consent between two sets of players. Therefore, the matching game can be expressed as a tuple,
\begin{equation}\label{game_tuple}
   {\cal G}\left( {{\cal S},{\cal C},{ \succ _n},{ \succ _l},{q_n}} \right).
\end{equation}

In classic matching games, quotas of two sets of players are both fixed. However, quotas of RVs in this network are unfixed, where one RV can serve dynamic number of SVs without substitutability. The matching problems of dynamic quotas and non-substitution motivate us to develop a new scheme that significantly differs from existing applications of matching game in wireless networks such as \cite{matching2,matching3,matching4}. Therefore, we propose a suitable solution to solve the problem of satisfaction-aware many-to-many relay assignment in (\ref{global_objective}).

\subsection{Proposed matching game without substitutability model}
In this paper, the many-to-many matching game without substitutability is proposed to optimize the global satisfaction. In the proposed game, SVs and relay radios select elements in the opposing set according to their own selection criteria.

\subsubsection{For SVs preferences}
In UAV communication networks with heterogeneous types of service, each SV $s_n$ seeks to find relay radios with high data rate so as to achieve its data rate requirement.
\begin{equation}\label{sourceprefer}
   {f_n}\left( {{\psi _n},{\psi _{ - n}}} \right) \to 1 \Rightarrow \text{find}{\rm{~}}\mu \left( {{s_n}} \right) \in \left\{ {{c_l}|{c_l} \in {\cal C}} \right\},\
\end{equation}
\begin{multline}\label{subjectsouce}
\begin{array}{l}
{\rm{s}}{\rm{.t}}{\rm{. }}~~~{u_n} = \sum\nolimits_{{c_l} \in {\cal C}} {\frac{{{C_{AF}}\left( {{s_n},{r_m}\left( {{c_l}} \right),d} \right){\delta _{nl}}}}{{A\left( {{c_l}} \right)}} \ge } {\rm{ }}{{u'}_n}\\
\\
{\rm{~~~~~~}}\sum\nolimits_{{c_l} \in {\cal C}} {{\delta _{nl}} \le {\alpha _n}}
\end{array}\
\end{multline}
where ${{f}_{n}}\left( {{\psi }_{n}},{{\psi }_{-n}} \right)\to 1$ means that the objective of $s_n$ is to obtain the desired transmission rate ${{{u}'}_{n}}$. Therefore, $s_n$ will search for suitable RVs to avoid the blind pursuit of high throughput.

According to throughput requirements, the preference of each SV ${{s}_{n}}$ can be expressed as,
\begin{equation}\label{preferlist}
    \left( {{s_n},{c_l}} \right){ \succ _n}\left( {{s_n},{c_j}} \right) \Leftrightarrow {u_n}\left( {{s_n},{c_l}} \right) > {u_n}\left( {{s_n},{c_j}} \right),
\end{equation}
where ${{u}_{n}}\left( {{s}_{n}},{{c}_{l}} \right)$ and ${{u}_{n}}\left( {{s}_{n}},{{c}_{j}} \right)$ represent the obtained throughput of ${{s}_{n}}$ assisted by ${{c}_{l}}$ and ${{c}_{j}}$, respectively. Therefore, relay radios which can provide better services will obtain higher priorities. It can be noted that subsets of source-relay pairs which can meet the requirement of $s_n$ have the same priority. The available resource of $s_n$ is impacted by the other SVs' choice dynamically, in which the situation belongs to one of the ``peer effects" \cite{multithree} in matching game model. Moreover, strategies of the other SVs are uncertain for $s_n$. Therefore, during the matching process, the preference ordering of one SV is dynamic according to the realistic data rate.

\subsubsection{Preferences of radios of RVs}
RVs assist the transmission of SVs to improve the global transmission performance. In the matching process, SVs propose matching proposals and then RVs decide whether to accept them according to the preference criterion. Because the selection strategy of one SV may influence the other SVs which have the same strategy sets, we define the utility function of relay radios as,
\begin{equation}\label{potentialcoming}
   \begin{array}{l}
{U_c}\left( {{s_n},{\psi _n}} \right) = {f_n}\left( {{\psi _n},{\psi _{ - n}}} \right) + \\
{\rm{~~~~~~~~~~~~~~~~~}}\sum\limits_{k \in {{\cal J}_n}} {\left[ {{f_k}\left( {{\psi _k},{\psi _{{{\cal J}_k}}}} \right) - {f_k}\left( {{\psi _k},{\psi _{{{\cal J}_k}\backslash n}}} \right)} \right]},
\end{array}\
\end{equation}
where $k\in {{\mathcal{J}}_{n}}$ denotes SVs which have the same relay preference elements in their preference lists, and they may be impacted by the decision of SV ${{s}_{n}}$. Therefore, ${{f}_{k}}\left( {{\psi }_{k}},{{\psi }_{{{\mathcal{J}}_{k}}}} \right)$ is the satisfaction result of the other SVs ${{s}_{k}}$. ${{f}_{k}}\left( {{\psi }_{k}},{{\psi }_{{{\mathcal{J}}_{k}}\backslash n}} \right)$ is the satisfaction result of ${{s}_{k}}$ if ${{s}_{n}}$ gives up the competition for radios.

It should be noted that not all the SVs would choose exactly the same radios even if the preference lists have overlapping parts. Some SVs that may cause interference to ${{s}_{n}}$ can be represented as
\begin{equation}\label{localcoming}
   {{\cal I}_n} = \left\{ {\forall {s_k}|{\delta _{kl}} = 1} \right\},{\rm{~~s}}{\rm{.t}}{\rm{.~~}}{c_l} \in \left\{ {{{\bar \psi }_n} \cup {\psi _n}} \right\}
\end{equation}
where ${{\bar \psi }_n}$ and ${\psi _n}$ mean the prepared selection strategy, and the current selection strategy of $s_n$, respectively. We assume that the selection strategy of $s_n$ changes from $\psi =\left\{ \psi \left| \psi \in \mathcal{C} \right. \right\}$ to $\bar{\psi }=\left\{ \bar{\psi }\left| \bar{\psi }\in \mathcal{C} \right. \right\}$. Only SVs in ${{\mathcal{I}}_{n}}$ can influence the transmission performance of ${{s}_{n}}$. Thus, it can be found that
\begin{equation}\label{changetolocal}
    {f_k}\left( {{\psi _k},{\psi _{{{\cal J}_k}}}} \right){\rm{ = }}{f_k}\left( {{\psi _k},{\psi _{{{\cal I}_k}}}} \right).\
\end{equation}
In this case, the satisfaction of SVs in ${{\cal J}_n}\backslash {{\cal I}_n}$ will not be influenced no matter which relay radios are chosen by $s_n$. Thus, it can also be found that
\begin{equation}\label{deletesome}
  {f_k}\left( {{\psi _k},{\psi _{{{\cal J}_k}}}} \right) = {f_k}\left( {{\psi _k},{\psi _{{{\cal J}_k}\backslash n}}} \right),k \in {{\cal J}_n}\backslash {{\cal I}_n}
\end{equation}

Based on (\ref{changetolocal}) and (\ref{deletesome}), ${{U}_{c}}\left( {{s}_{n}},{{\psi }_{n}} \right)$ can be simplified to
\begin{equation}\label{simplified}
   \begin{array}{l}
{U_c}\left( {{s_n},{\psi _n}} \right) = {f_n}\left( {{\psi _n},{\psi _{ - n}}} \right) + \\
~~~~~~~~\sum\limits_{k \in {{\cal I}_n}} {\left[ {{f_k}\left( {{\psi _k},{\psi _{{{\cal I}_k}}}} \right) - {f_k}\left( {{\psi _k},{\psi _{{{\cal I}_k}\backslash n}}} \right)} \right]}.
\end{array}\
\end{equation}
Relay radios controlled by RVs determine whether to accept strategies changes of $s_n$. Different from the classic matching process, matching processes in the proposed many-to-many matching game without substitutability are not substituted. SVs share resources of relay radios rather than replace each other. Inspired by the potential function in \cite{potentialgame}, the optimal solution of the satisfaction maximization problem in (\ref{simplified}) will lead to a global stable matching result.
\begin{definition}\label{NEquilibrium}
A matching results is a global stable matching if and only if no player can improve its matching utility by deviating unilaterally, i.e.,
\begin{equation}\label{NEq}
\nexists\Lambda \left( {{{\bar \psi }_n},{\psi _{ - n}}} \right) > \Lambda \left( {{\psi _n},{\psi _{ - n}}} \right),
\end{equation}
\begin{equation}\label{subjectto}
    {\rm{s}}{\rm{.t}}{\rm{.~~~}}\forall {s_n} \in {\cal N},{\rm{   }}{\psi _n} \subseteq {\cal C},{\rm{   }}\left| {{\psi _n}} \right| \le {\alpha _n}~~~~~\
\end{equation}

\end{definition}

\begin{theorem}\label{potenialtoNE}
Optimizing the local matching process, the global network will achieve a stable matching result.
\end{theorem}

\begin{IEEEproof}
Due to space limitations, this article does not give a specific proof. The specific proof will be given in \cite{futurework}.
\end{IEEEproof}

It is shown that the many-to-many game model without substitutability has at least one stable matching result. In this game model, SVs needn't exchange information with the other SVs, and RVs only need to exchange information with a portion of the other RVs in the same strategy sets of SVs. Therefore, the game model reduces the information exchanges of SVs as well as RVs.

\section{Potential matching approach for many-to-many matching market}
In order to study the relay selection with limited information exchanges, a distributed potential matching approach (PMA) is proposed in this paper, in which each SV knows the channel state information (CSI) so as to reorganize preference ordering of relay radios.

\begin{table}[t!]
\emph{\textbf{Algorithm1}: Distributed potential matching approach (PMA) for many-to-many relay selection network}
\\
\rule{\linewidth}{1pt}
\textbf{Initialization:} Each SV randomly chooses relay radios;

\textbf{Loop} in iteration $k$;

\textbf{Stage I:} Relay radios estimate and utilities computation

~~Each SV calculates utility function over its all available relay radio, then it chooses a set of available relay radios based on the mixed strategy, where the component that denotes the probability of radio selection $c_l$ in the mixed strategy is given as

\begin{equation}\label{NEXTselection}
    {{\bar \psi }_n}\left( {c_l} \right) = \frac{{u\left( {{s_n},{c_l}} \right)}}{{\sum\nolimits_{{c_l} \in {\cal C}} {u\left( {{s_n},{c_l}} \right)} }}.
\end{equation}

\textbf{Stage II:} Matching evaluation

~~Based on (\ref{simplified}), relative RVs decide if accept or not with a mixed strategy, where the component contains ``accept" or ``reject".

\begin{equation}\label{mixstrategies}
    P = \frac{{\exp \left\{ {\beta  \cdot {U_c}\left( {{{\bar \psi }_n}} \right),\beta  \cdot {U_c}\left( {{\psi _n}} \right)} \right\}}}{{\exp \left\{ {\beta  \cdot {U_c}\left( {{{\bar \psi }_n}} \right)} \right\} + \exp \left\{ {\beta  \cdot {U_c}\left( {{\psi _n}} \right)} \right\}}},
\end{equation}
for some learning parameter $\beta  > 0$.

\textbf{end Loop} until $\not \exists \mu _{i,j}^m$ can improve the satisfaction results.

\textbf{Output:} Convergence to a stable matching result.

\rule{\linewidth}{1pt}
\end{table}

Here, PMA with limited information exchanges is proposed in Algorithm 1, which consists of two stages: utility computation of relay radio and matching evaluation. In Stage I, based on the location information in the relay system, each SV forms its own preference list according to the CSI. Although SVs obtain the CSI information in every iteration, they can obtain neither strategies of the other SVs, nor influences of their own strategies. Therefore, it randomly chooses a set of relay radios according to the mixed strategy (\ref{NEXTselection}), where the number of connection is no more than the number of its own radios. SVs ask requests of connection exchange to RVs, and RVs will accept the exchange proposal according to (\ref{mixstrategies}). With a sufficiently large $\beta$, the global satisfaction maximization can be achieved with an arbitrarily high probability \cite{largefactor}.

Note that RVs receive proposals from SVs and only need to exchange information with the relative RVs. The approach decreases the information exchanges and mitigates the costs of SVs and the global network.

\section{Simulation Results and Discussions}\label{SEC:5Simulation}
In a 2000$\times$2000 square metre topology, there is a UAV controller located at centre of this topology. Several RVs are close to the UAV controller, and cell edged SVs are randomly distributed. Each UAV is equipped with two radios, and radios of RVs work in orthogonal channels. SVs transmit data to the destination node respectively, and the experimental parameters follow the simulation methodologies of 3GPP specifications \cite{3GPP}.

Assume that $W$ = 10 MHz bandwidth for each channel of the system, and the noise power density of the system is -174 dBm/Hz. The maximum transmission powers of SVs are set to 20 dBm and RVs are set to 30 dBm. SVs have individual data rate requirements range from 10 Mbps to 40 Mbps randomly and the $\alpha $ of each SV also is random ($0< \alpha \le 3$). All results are obtained by simulating 600 topologies independently and taking the expected values. To balance the tradeoff between exploration and exploitation, $\beta=k$ is chosen in our simulation, where $k$ means the the iteration step.
\subsection{Convergence performance}
We consider a UAV communication network consisting of five RVs and thirteen SVs. Therefore, thirteen SVs share ten relay radios according to different requirements. It can be seen that in this scenario, the maximum number of possible strategy selection profiles is ${\left( {{\rm{C}}_{10}^2 + 10} \right)^{13}}$.

\begin{figure}
  \includegraphics[width=3in]{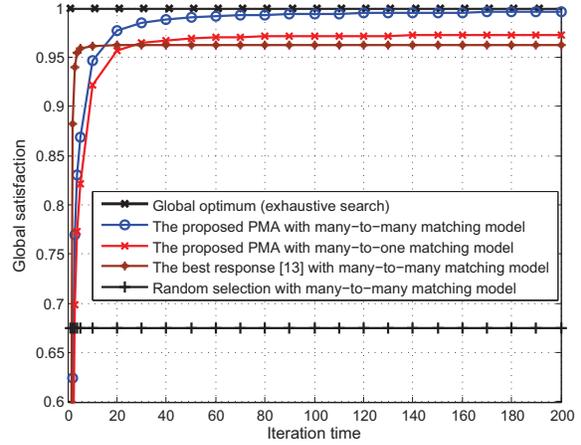}\\
  \caption{Evolution results of satisfaction performance with 13 SVs and 10 relay radios.}\label{contobest}
\end{figure}

The average convergence behavior of five approaches are shown in Fig. \ref{contobest}, in which the global optimum is obtained by using the exhaustive search method. It is noted from the figure that the proposed distributed PMA with many-to-many matching model catches up with the global optimum. Also, the performance comparison of many-to-one relay selection (each SV chooses one radio at most) and the best response algorithm \cite{potentialgame} with many-to-many matching model is given. Both two algorithms can not achieve the optimal performance. It can be noted that the result of many-to-many selection is better than that of the many-to-one selection, which means that the many-to-many relay selection can achieve more flexible resource assignment in the network with heterogeneous requirements.

\begin{figure}
  \includegraphics[width=3in]{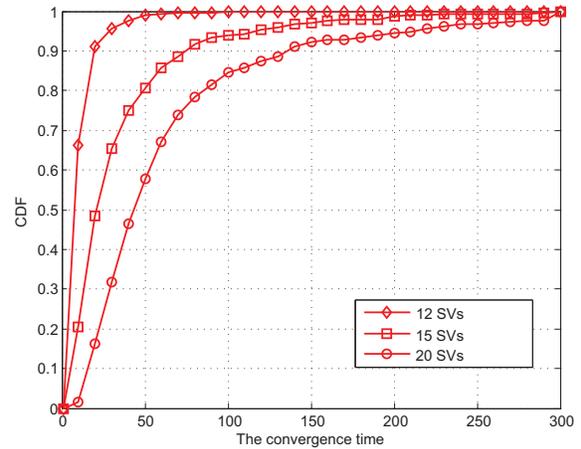}\\
  \caption{The performance of convergence with kinds of numbers of SVs in the network.}\label{convergencesource}
\end{figure}

Fig. \ref{convergencesource} represents the cumulative distribution function (CDF) for the convergence time of the proposed algorithm. In Fig. \ref{convergencesource}, we can see that, the average number of iteration increases due to the increase of the number of players in the UAV network. With the number of devices increases, the collisions among SVs also increase. However, the collisions have a final upper boundary because the UAV network becomes a saturated state. Fig. \ref{convergencesource} demonstrates that the proposed matching approach has a reasonable convergence time that does not exceed 300 iterations with 20 SVs and 10 relay radios in the incomplete information network.

\subsection{Satisfaction performance comparison}
\begin{figure}
  \includegraphics[width=3in]{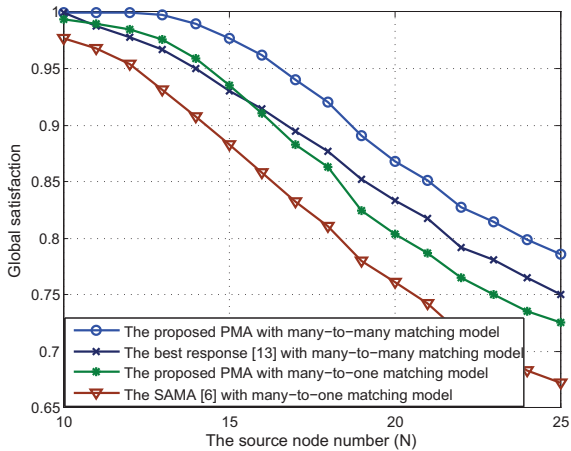}\\
  \caption{The comparison of average satisfaction among SVs with heterogeneous requirements.}\label{SATISFAC}
\end{figure}

The average satisfaction performance of different algorithms with varying network scales are compared. We consider relay networks with five RVs and the number of SVs are varying. Fig. \ref{SATISFAC} shows the proportion value of the average satisfaction resulting by the proposed PMA with many-to-many model, PMA with many-to-one matching model, the best response algorithm \cite{potentialgame} with many-to-many matching model and SAMA \cite{multithree} with many-to-one matching model.

It is illustrated in Fig. \ref{SATISFAC} that, the proposed PMA has a significant advantage in term of satisfaction at all network sizes and the proportion of satisfaction is higher than 0.95 when the number of source-destination pairs is not more than 16. It can be noted that, the best response algorithm \cite{potentialgame} is better than the PMA with many-to-one matching model when the number of SVs is higher than 15, while the contrary result is obtained when the number of SVs less than 15. The SAMA \cite{multithree} is a matching algorithm with substitutability, and the performance is worse than that of the PMA. These results mean that 1) many-to-many selection strategies can obtain more flexible resource allocation in heterogeneous requirement networks; 2) unsuitable many-to-many selection strategies may achieve worse results than that of many-to-one relay model; 3) the relay selection network can achieve better performances by the matching game without substitutability.

\section{Conclusion}\label{SEC:6Conclusion}
In this paper, the selection problem of the multi-channel multi-radio UAV relay network has been studied. We have formulated the selection network as a many-to-many matching market. To solve the selection problem, we have proposed a many-to-many matching game without substitutability. A distributed potential matching algorithm has been designed, in which the local matching processes lead to a global stable matching result. Simulation results have shown that the proposed potential-matching approach yields stable matching results, which outperform existing schemes with the objective of satisfaction optimization. In UAV networks with heterogeneous requirements, many-to-many approach achieves more flexible resource assignments. Based on this work, the transmission characteristics of dynamic UAV communication networks will be studied in the future.

\section*{Acknowledgement}
This work was supported by the National Science Foundation of China under Grant No. 61771488, No. 61671473, No. 61631020 and No. 61401508, the in part by Natural Science Foundation for Distinguished Young Scholars of Jiangsu Province under Grant No. BK20160034, and in part by the Open Research Foundation of Science and Technology on Communication Networks Laboratory. The authors would like to thank Miss Mengyun Tang for her helpful discussions.

\end{document}